\documentclass[review]{elsarticle}
\usepackage[letterpaper,top=2cm,bottom=2cm,left=3cm,right=3cm,marginparwidth=1.75cm]{geometry}
\usepackage{lineno,hyperref,mathtools}

\journal{Journal of \LaTeX\ Templates}

\usepackage{amssymb}
\usepackage{bm}
\usepackage{amssymb}
\usepackage{amsthm}
\usepackage{multirow,booktabs}
\usepackage{cases}
\usepackage{threeparttable}
\usepackage{color}
\usepackage{rotating}
\newtheorem{theorem}{Theorem}

\newtheorem{remark}{Remark}
\newtheorem{definition}{Definition}
\newtheorem{lemma}{Lemma}
\newtheorem{corollary}{Corollary}
\newtheorem{example}{Example}

\newcommand{\Hull}{{\mathrm{Hull}}}
\newcommand{\C}{{\mathcal{C}}}
\newcommand{\F}{{\mathbb{F}}}

\newcommand{\aaa}{{\bm{\alpha}}}
\newcommand{\as}{\alpha}

\begin{document}

\begin{frontmatter}

\title{On the Existence of Galois Self-Dual GRS and TGRS Codes}
\tnotetext[mytitlenote]{This research work is supported by the National Natural Science Foundation of China under Grant Nos. U21A20428 and 12171134.}

\author[mymainaddress]{Shixin Zhu}
\ead{zhushixinmath@hfut.edu.cn}

\author[mymainaddress]{Ruhao Wan\corref{mycorrespondingauthor}}
\cortext[mycorrespondingauthor]{Corresponding author}
\ead{wanruhao98@163.com}

\address[mymainaddress]{School of Mathematics, HeFei University of Technology, Hefei 230601, China}

\begin{abstract}
Let $q=p^m$ be a prime power and $e$ be an integer
with $0\leq e\leq m-1$.
$e$-Galois self-dual codes are generalizations
of Euclidean $(e=0)$ and Hermitian ($e=\frac{m}{2}$ with even $m$)
self-dual codes.
In this paper,
for a linear code $\C$ and a nonzero vector $\bm{u}\in \F_q^n$, we give a sufficient and necessary condition for the dual extended code $\underline{\C}[\bm{u}]$ of $\C$ to be $e$-Galois self-orthogonal.
From this, a new systematic approach is proposed to prove the existence of $e$-Galois self-dual codes.
By this method, we prove that $e$-Galois self-dual (extended) generalized Reed-Solomon (GRS) codes of length $n>\min\{p^e+1,p^{m-e}+1\}$ do not
exist, where $1\leq e\leq m-1$.
Moreover, based on the non-GRS properties of twisted GRS (TGRS) codes, we show that in many cases $e$-Galois self-dual (extended) TGRS codes do not exist.
Furthermore,
we present a sufficient and necessary condition for $(\ast)$-TGRS codes to be Hermitian self-dual,
and then construct several new classes of Hermitian self-dual $(+)$-TGRS and $(\ast)$-TGRS codes.
\end{abstract}


\end{frontmatter}

\section{Introduction}\label{sec1}

Throughout this paper, $q=p^m$ is a prime power and $0\leq e\leq m-1$ is an integer.
Let $\F_{q}$ be a finite field with $q$ elements and $\F_q^*=\F_q\backslash \{0\}$.
An $[n,k,d]_{q}$ linear code $\C$ is a subspace of $\F_{q}^{n}$ with dimension $k$ and minimum distance $d$.
For a linear code $\C$, it must satisfy the \emph{Singleton bound}: $d\leq n-k+1$.
If $d=n-k+1$, $\C$ is called an \emph{maximum distance separable} (MDS) \emph{code}.
If $d=n-k$, then $\C$ is called \emph{almost} MDS (AMDS).
In addition, $\C$ is said to be \emph{near} MDS (NMDS) if both $\C$ and $\C^{\bot_E}$ are AMDS,
where $\C^{\bot_E}$ is the Euclidean dual code of $\C$.
Due to the nice algebraic structure and error correction capability of MDS and NMDS codes, they are important in coding theory and have a wide range of applications
(see \cite{RefJ (2003) W. Fund}-\cite{RefJ (1994) Dougherty NMDS}).
Then the study of MDS and NMDS codes has attracted a lot of attention (see \cite{RefJ (2008) M.G}-\cite{RefJ10}).
Particularly, GRS codes and \emph{extended} GRS (EGRS) codes, as equivalent classes of codes ($n\leq q$) (see \cite{RefJ (1996) Pellikaan}), are the most important families of MDS codes.
A lot of MDS self-dual codes are constructed based on GRS and EGRS codes (see \cite{RefJ12}-\cite{RefJ18} and the references therein).

 In \cite{RefJ (2017) Bee TGRS}, Beelen et al. first introduced \emph{twisted} GRS (TGRS) codes, which is a generalization of GRS codes.
Different from GRS codes, they showed that TGRS codes are not necessarily MDS and presented a sufficient and necessary condition for TGRS codes to be MDS (see \cite{RefJ (2017) Bee TGRS, RefJ (2022) Bee TGRS}).
Based on the non-GRS properties of TGRS codes, TGRS codes are resistant to Sidelnikov-Shestakov attacks and Wieschebrink attacks, whereas GRS codes are not (see \cite{RefJ (2018) Bee TGRS,RefJ (2020) Lav track}).
For this reason, the construction of self-dual TGRS codes has received much attention in recent years and some important processes have been made in the study of (extended) TGRS codes (see \cite{RefJ (2017) Bee TGRS}-\cite{RefJ (2024) Zhu (+)} and the references therein).

  In \cite{RefJ (2017) Galois}, Fan et al. first introduced the Galois inner product.
  Since then, the Galois inner product has attracted much attention as a generalisation of the Euclidean and Hermitian inner product.
  Specifically, in \cite{RefJ (2017) Galois}-\cite{RefJ (2023) Fu DM}, sufficient
conditions (some are also necessary) for (extended) constacyclic codes and skew multi-twisted codes
over $\F_q$ to be Galois self-orthogonal or Galois self-dual were presented.
The $e$-Galois hull of a linear code $\C$ is defined to be $\Hull_e(\C)=\C\cap \C^{\bot_e}$, where $\C^{\bot_e}$ is the $e$-Galois dual code of $\C$.
Due to the excellent properties of the hull of linear codes, some researchers began to study the Galois hull of linear codes (see \cite{RefJ (2020) Liu. Galois,RefJ (2023) Liu Galois}).
The results on Galois hulls of linear codes have
important applications in the constructions of entanglement-assisted quantum error-correcting codes (EAQECCs).
In particular, in \cite{RefJ (2019) Liu  FFA, RefJ (2020) Liu  QIP}, Liu et al. constructed several classes of EAQECCs via Galois dual codes or Galois LCD codes.
Recently,
with the aims of constructing EAQECCs and MDS codes with Galois hulls of arbitrary
dimensions, GRS codes have been studied under the Galois inner product
(see \cite{RefJ (2021) cao}-\cite{RefJ (2024) Qian galois} and the references therein).

\subsection{Our motivation}

Our main motivations can be summarized as follows:

\begin{itemize}
\item

\begin{itemize}
\item
On the one hand, giving conditions for the existence of Galois self-dual codes has received much attention in recent years.
For example,
in \cite{RefJ (2017) Galois}, Fan et al. gave existence conditions of $e$-Galois self-dual constacyclic codes.
In \cite{RefJ (2021) Mi DM}, Mi et al. constructed all normal MDS $e$-Galois self-dual
constacyclic codes.
In \cite{RefJ (2023) Fu DM},
Fu et al. gave existence
conditions of Galois self-dual codes which are extensions of constacyclic codes.

\item
On the other hand,
compared to the Euclidean and Hermitian inner product, the Galois inner product
has the more general setting, which allows us to to find more codes with better algebraic structures and good parameters.
For example, in \cite{RefJ (2021) cao}-\cite{RefJ (2024) Qian galois}, the dimension of the constructed MDS codes with $e$-Galois hulls is often related to $e$.

\end{itemize}

 Hence, it is a valuable work to give a systematic approach to obtain the existence of Galois self-dual codes.
 Once the existence of Galois self-dual codes is available, the existence of the
associated Euclidean and Hermitian self-dual codes can be
obtained immediately by special cases.

\item
In \cite{RefJ (2022) Ball determine}, Ball et al. transformed the existence of an Hermitian self-orthogonal GRS code into the
existence of a polynomial with a given number of distinct zeros,
 and proved Conjecture 11 in \cite{RefJ (2015) n=q^2+1(2)}.
From this we know that there exists no $q^2$-ary Hermitian self-dual GRS code, for even length $n>q+1$.
Naturally, a question in this topic is: for more general Galois inner products, what about the existence of Galois self-dual GRS codes?

\item
 In recent years, (extended) TGRS codes as
a generalisation of (extended) GRS codes have been widely used to construct self-dual
(non-GRS) MDS or NMDS codes.
Naturally, a question in this topic is: which (extended) TGRS codes can be self-dual and which cannot?
Therefore, it is an interesting work to give the existence of
self-dual TGRS codes. The same is true for the Galois inner product as a generalisation.

\item
In \cite{RefJ2}-\cite{RefJ1}, the authors constructed $q^2$-ary Hermitian self-dual MDS codes of even length $n\leq q+1$ from (extended) GRS codes.
Recently, in \cite{RefJ (2023) Guo H-TGRS}, Guo et al. gave
a sufficient and necessary condition for $(+)$-TGRS codes to be Hermitian self-dual and constructed two
classes of Hermitian self-dual $(+)$-TGRS codes.
Note that non-GRS codes are Sidelnikov-Shestakov attacks and Wieschebrink attack resistent (see \cite{RefJ (2018) Bee TGRS,RefJ (2020) Lav track}),
then it makes sense to construct more Hermitian self-dual non-GRS MDS or NMDS codes by TGRS codes.
\end{itemize}

\subsection{Our results}

Recently, for a given linear code $\C$ and a nonzero vector $\bm{u}\in \F_q^n$, Sun et al. \cite{RefJ (2023) Sun extended} defined an extended linear code $\underline{\C}(\bm{u})$ of $\C$, which is a generalization of the classical extended code.
In this paper, a new systematic approach is proposed to prove the existence of Galois self-dual codes by means of extended codes of linear codes. By applying the new method,
some non-existence results of Galois self-dual GRS and TGRS codes are obtained.
Moreover, several new classes of Hermitian self-dual $(+)$-TGRS and $(\ast)$-TGRS codes are constructed.
The main contributions of this paper can be summarized as follows:

\begin{itemize}
\item
For a linear code $\C$ and a nonzero vector $\bm{u}\in \F_q^n$, we give a sufficient and necessary condition for the dual extended code $\underline{\C}[\bm{u}]$ of $\C$ (see Definition \ref{def E[]}) to be $e$-Galois self-orthogonal (see Theorem \ref{th galois 1}).
From this we can directly obtain a sufficient and necessary condition for the extended code $\underline{\C}(\bm{u})$ of $\C$
(see Definition \ref{def E()}) to be $e$-Galois dual-containing (see Corollary \ref{cor galois 1}).

\item
Sufficient and necessary conditions for EGRS codes to be $e$-Galois self-dual are
presented (see Theorem \ref{th galois EGRS}).
From this we prove that $e$-Galois self-dual (extended) GRS codes of length $n>\min\{p^e+1,p^{m-e}+1\}$ do not exist, where $1\leq e\leq m-1$ (see Theorem \ref{th GRS not exist}).

\item
Sufficient and necessary conditions for TGRS codes with $h=0$ and $0\in A_{\aaa}$
 (resp. ETGRS codes with $h=k-1$) to be $e$-Galois self-dual are presented (see Theorems \ref{th galois *TGRS} and \ref{th galois ETGRS}).
Then based on the non-GRS properties of TGRS codes,
we show that in many cases $e$-Galois self-dual TGRS and ETGRS codes do not exist (see Theorems \ref{th TGRS not exist} and \ref{th ETGRS not exist}).

\item
Finally, sufficient and necessary conditions for $(\ast)$-TGRS codes to be Hermitian self-dual are presented
(see Lemma \ref{lem H-dual *TGRS}).
From this, we construct several new classes of Hermitian self-dual $(+)$-TGRS and $(\ast)$-TGRS codes
(see Theorems \ref{th +TGRS} and \ref{th *TGRS}),
 the parameters of these Hermitian self-dual codes can be different from those Hermitian self-dual $(+)$-TGRS codes in \cite{RefJ (2023) Guo H-TGRS}.
\end{itemize}

\subsection{Organization of this paper}

The rest of this paper is organized as follows.
In Section \ref{sec2}, we briefly introduce some basic notations and results on (dual) extended codes and Galois self-orthogonal codes, and give a sufficient and necessary condition for dual extended codes to be Galois self-orthogonal.
In Section \ref{sec3}, we present our main results on the existence of Galois self-dual GRS and TGRS codes.
In Section \ref{sec4}, several new classes of Hermitian self-dual $(+)$-TGRS and $(\ast)$-TGRS codes are constructed.
Finally, we give a short summary of this paper in Section \ref{sec5}.

\section{Galois self-orthogonal dual extended codes}\label{sec2}

In this section, we introduce some basic results about Galois self-orthogonal and give a sufficient and necessary condition for dual extended codes to be Galois self-orthogonal.

Let $q=p^m$, where $p$ is prime and $0\leq e\leq m-1$ be an integer.
Let $\F_q$ be the finite field with $q$ elements and $\F_{q}^*=\F_{q}\backslash \{0\}$.
An $[n,k,d]_{q}$ linear code $\C$ over $\F_{q}$ can be seen as a $k$-dimensional subspace of $\F_{q}^n$ with minimum distance $d$.
Suppose that $\bm{x}=(x_1,x_2,\dots,x_n)$ and $\bm{y}=(y_1,y_2,\dots,y_n)$
are two vectors in $\F_{q}^n$,
then the \emph{$e$-Galois inner product} of vectors $\bm{x}$ and $\bm{y}$ is defined as
$$\langle \bm{x},\bm{y} \rangle_{e}=\sum_{i=1}^{n}x_iy_i^{p^e},$$
where $e$ is an integer with $0\leq e\leq m-1$.
The $e$-Galois inner product is a generalization of the \emph{Euclidean inner product} (i.e., $e=0$) and the \emph{Hermitian inner product} (i.e., $e=\frac{m}{2}$ with even $m$).
For convenience, we use $\langle - ,- \rangle_E$ (resp. $\langle -,- \rangle_H$) to denote $\langle -, -\rangle_{0}$ (resp. $\langle -,-\rangle_{\frac{m}{2}}$ with even $m$).
The \emph{$e$-Galois dual code} of $\C$ is defined as
$$\C^{\perp_{e}}=\Big \{\bm{y}\in \F_{q}^n:\langle \bm{x}, \bm{y}\rangle_{e}=0, \ {\rm for \ all}\ \bm{x}\in \C \Big \}.$$
Then $\C^{\bot_E}=\C^{\bot_0}$ (resp. $\C^{\bot_H}=\C^{\bot_{\frac{m}{2}}}$ with even $m$) is just the \emph{Euclidean} (resp. \emph{Hermitian}) \emph{dual code} of $\C$.
In particular, $\C$ is called \emph{$e$-Galois self-orthogonal}
if $\C\subseteq \C^{\bot_e}$, \emph{$e$-Galois dual-containing} if $\C^{\bot_{e}}\subseteq \C$ and \emph{$e$-Galois self-dual} if $\C=\C^{\bot_{e}}$.
We fix some notations as follows for convenience.
\begin{itemize}
\item For a vector $\aaa=(\as_1,\as_2,\dots,\as_n)\in \F_{q}^n$, with $\as_i\neq \as_j\ (i\neq j)$,
denote
$$A_{\aaa}=\{\as_1,\as_2,\dots, \as_n\},\ L_{\aaa}=(L_{\aaa}(\as_1),L_{\aaa}(\as_2),\dots,L_{\aaa}(\as_n))\ {\rm and} \ \aaa^z=(\as_1^z,\as_2^z,\dots,\as_n^z),$$
 where $L_{\aaa}(\as_i)=\prod_{1\leq j\leq n,j\neq i}(\as_i-\as_j)$ and $z$ is an integer.
Specially, $0^0=1$. Set $s(\aaa)=\sum_{i=1}^n \as_i$.
\item
Let $\bm{0}$ (resp. $\bm{1}$) be the all zero (resp. one) vector and the length of $\bm{0}$ (resp. $\bm{1}$) depends on the context.
$\bm{0}_{k\times n}$ denotes the $k\times n$ zero matrix.
\item
For $k\times n$ matrix $G$ over $\F_q$ with row vectors $\bm{g}_1,\bm{g}_2,\dots,\bm{g}_k\in \F_q^n$, write $G=(\bm{g}_1,\bm{g}_2,\dots,\bm{g}_k)^{\mathcal{T}}$.
\item
For $\bm{x}=(x_1,x_2,\dots,x_n)\in \F_{q}^n$ and $\bm{y}=(y_1,y_2,\dots,y_n)\in \F_{q}^n$,
the \emph{Schur product} between $\bm{x}$ and $\bm{y}$ is defined as
$$\bm{x}\star \bm{y}=(x_1y_1,x_2y_2,\dots,x_ny_n).$$
\item Let $\sigma$: $\F_q\rightarrow\F_q$, $a\mapsto a^p$ be the \emph{Frobenius automorphism} of $\F_q$.
For any vector $\bm{x}=(x_1,x_2,\dots,x_n)\in \F_q^n$ and any matrix $G=(g_{ij})_{k\times n}$ over $\F_q$,
we denote
$$\sigma(\bm{x})=(\sigma(x_1),\sigma(x_2),\dots,\sigma(x_n))\quad {\rm and}\quad \sigma(G)=(\sigma(g_{ij}))_{k\times n}.$$
Similarly, the mapping $\sigma^e$: $\F_q\rightarrow\F_q$, $\sigma^e(a)=a^{p^e}$, $\forall\ a\in \F_q$
is an automorphism of $\F_q$.
Moreover, the inverse of the mapping
$\sigma^e$ is denoted by $\sigma^{m-e}$: $\sigma^{m-e}(a)=a^{p^{m-e}}$.
\end{itemize}

There are some important results on Galois dual codes in the literature.
We review them here.

\begin{lemma}\label{lem galois OA}
(\cite{RefJ (2018) Liu. Galois} and \cite{RefJ (2020) Liu. Galois})
Let $\C$ be an $[n,k]_q$ linear code with a generator matrix $G$.
Then $\C$ is an $e$-Galois self-orthogonal code if and only if $\C$ is an $(m-e)$-Galois self-orthogonal code,
if and only if $G\sigma^e(G^T)=\bm{0}_{k\times k}$, if and only if $G\sigma^{m-e}(G^T)=\bm{0}_{k\times k}$.
\end{lemma}

\begin{lemma}\label{lem galois e}
(\cite{RefJ (2018) Liu. Galois} and \cite{RefJ (2020) Liu. Galois})
Let $\C$ be an $[n,k]_q$ linear code with a generator matrix $G$. Then $\sigma^{m-e}(\C)$
is an $[n,k]_q$ linear code with a generator matrix $\sigma^{m-e}(G)$ and $\C^{\bot_e}=(\sigma^{m-e}(\C))^{\bot_E}=\sigma^{m-e}(\C^{\bot_E})$.
Moreover, $(\C^{\bot_e})^{\bot_{m-e}}=\C$.
\end{lemma}

For a linear code $\C$ of length $n$ and a nonzero vector $\bm{u}\in \F_q^n$, we can define the \emph{extended code} of $\C$ as follows.

\begin{definition}\label{def E()}
(\cite{RefJ (2023) Sun extended})
Let $\bm{u}=(u_1,u_2,\dots,u_n)\in \F_q^n$ be any nonzero vector.
For a given $[n,k,d]_q$ linear code $\C$,
define an $[n+1,k,\bar{d}]_q$ code $\underline{\C}(\bm{u})$ by
$$\underline{\C}(\bm{u})=\Big \{(c_1,\dots,c_n,c_{n+1}):\ (c_1,c_2,\dots,c_n)\in\C,c_{n+1}=\sum_{i=1}^n u_ic_i \Big\},$$
where $\bar{d}=d$ or $\bar{d}=d+1$.
\end{definition}

It is easy to check that the following lemma holds by the definition of extended codes.

\begin{lemma}\label{lem E()}
(\cite{RefJ (2023) Sun extended})
Let $\C$ be an $[n,k]_q$ linear code with
generator matrix $G$ and parity check matrix $H$.
Let $\bm{u}\in \F_q^n$.
Then the generator and parity check matrices for the extended code $\underline{\C}(\bm{u})$ are
$$\underline{G}(\bm{u})\triangleq(G,G\bm{u}^T)\quad and\quad  \underline{H}[\bm{u}]\triangleq\begin{pmatrix}
H& \bm{0}^T\\
\bm{u} &-1\\
\end{pmatrix}.$$
\end{lemma}

By the extended code of $\C$, we define the \emph{dual extended code} of $\C$ as follows.

\begin{definition}\label{def E[]}
Let $\bm{u}\in \F_q^n$ be any nonzero vector.
For a given $[n,k]_q$ linear code $\C$,
define an $[n+1,k+1]_q$ code $\underline{\C}[\bm{u}]$ by
$$\underline{\C}[\bm{u}]=(\underline{\C^{\bot_E}}(\bm{u}))^{\bot_E}.$$
\end{definition}

Similarly, it is direct to obtain the following lemma.

\begin{lemma}\label{lem E[]}
Let $\C$ be an $[n,k]_q$ linear code with
generator matrix $G$ and parity check matrix $H$.
Let $\bm{u}\in \F_q^n$ be any nonzero vector.
Then the generator and parity check matrices for the dual extended code $\underline{\C}[\bm{u}]$ are
$$\underline{G}[\bm{u}]\triangleq\begin{pmatrix}
G& \bm{0}^T\\
\bm{u} &-1\\
\end{pmatrix}
\quad and\quad  \underline{H}(\bm{u})\triangleq(H,H\bm{u}^T).$$
\end{lemma}

We now present our key theorem of this paper, which gives a sufficient and necessary condition for $\underline{\C}[\bm{u}]$ to be
$e$-Galois self-orthogonal.

\begin{theorem}\label{th galois 1}
Let $\C$ be an $[n,k]_q$ linear code
and $\bm{u}\in \F_q^n$ be a nonzero vector.
Then $\underline{\C}[\bm{u}]$ is $e$-Galois self-orthogonal if and only if
$\C \subseteq \C^{\bot_e}$, $\bm{u}\in \C^{\bot_e}$,
$\sigma^{m-2e}(\bm{u})\in \C^{\bot_e}$
and $s(\bm{u}^{p^e+1})=(-1)^{p^e}$.
\end{theorem}

\begin{proof}
Suppose that $\C$ have a generator matrix $G$.
By Lemma \ref{lem E[]},
it follows that $\underline{\C}[\bm{u}]$ has a generator matrix $\underline{G}[\bm{u}]$.
Note that
\[\begin{split}
\underline{G}[\bm{u}]\sigma^e(\underline{G}[\bm{u}]^T)=&\begin{pmatrix}
G& \bm{0}^T\\
\bm{u} &-1\\
\end{pmatrix} \begin{pmatrix}
\sigma^e (G)& \bm{0}^T\\
\sigma^e(\bm{u}) & (-1)^{p^e}\\
\end{pmatrix}^T\\
=&\begin{pmatrix}
G& \bm{0}^T\\
\bm{u} &-1\\
\end{pmatrix} \begin{pmatrix}
\sigma^e (G^T)& \sigma^e(\bm{u}^T)\\
\bm{0} & (-1)^{p^e}\\
\end{pmatrix}\\
=&\begin{pmatrix}
G\sigma^e (G^T) & G\sigma^e(\bm{u}^T) \\
\bm{u}\sigma^e (G^T) & s(\bm{u}^{p^e+1})+(-1)^{p^e+1}\\
\end{pmatrix},\\
\end{split}\]
then by Lemma \ref{lem galois OA}, $\underline{\C}[\bm{u}]$ is $e$-Galois self-orthogonal if and only if $\underline{G}[\bm{u}]\sigma^e(\underline{G}[\bm{u}]^T)=\bm{0}_{(k+1)\times (k+1)}$
if and only if $G\sigma^e (G^T)=\bm{0}_{k\times k}$, $G\sigma^e(\bm{u}^T)=\bm{0}^T$, $\bm{u}\sigma^e (G^T)=\bm{0}$ and $s(\bm{u}^{p^e+1})=(-1)^{p^e}$.
Since $\sigma^{m-e}$ is the inverse map of $\sigma^e$, then we have
$$\bm{u}\sigma^e (G^T)=\bm{0}\Leftrightarrow  \sigma^{m-e}(\bm{u}) (G^T)=\bm{0}
 \Leftrightarrow  G\sigma^{m-e}(\bm{u}^T)=\bm{0}^T.$$
Note that
$$G\sigma^e(\bm{u}^T)=\bm{0}^T \Leftrightarrow \bm{u}\in \C^{\bot_e}\quad {\rm and}\quad  G\sigma^{m-e}(\bm{u}^T)=\bm{0}^T
 \Leftrightarrow  \sigma^{m-2e}(\bm{u})\in \C^{\bot_e},$$
 then by Lemma  \ref{lem galois OA}, $\underline{\C}[\bm{u}]$ is $e$-Galois self-orthogonal
if and only if
$\C \subseteq \C^{\bot_e}$, $\bm{u}\in \C^{\bot_e}$,
$\sigma^{m-2e}(\bm{u})\in \C^{\bot_e}$
and $s(\bm{u}^{p^e+1})=(-1)^{p^e}$.
This completes the proof.
\end{proof}

Note that $\underline{\C}[\bm{u}]^{\bot_E}=\underline{\C^{\bot_E}}(\bm{u})$, we can obtain the following corollary,
which gives a sufficient and necessary condition for $\underline{\C}(\bm{u})$ to be
$e$-Galois dual-containing.

\begin{corollary}\label{cor galois 1}
Let $\C$ be an $[n,k]_q$ linear code
and $\bm{u}\in \F_q^n$ be a nonzero vector.
Then $\underline{\C}(\bm{u})$ is $e$-Galois dual-containing if and only if
$\C^{\bot_{e}}\subseteq \C$, $\sigma^e(\bm{u})\in \C$,
$\sigma^{m-e}(\bm{u})\in \C$ and $s(\bm{u}^{p^e+1})=(-1)^{p^e}$.
\end{corollary}

\begin{proof}
By Lemmas \ref{lem galois OA} and \ref{lem galois e}, we have
 \[\begin{split}  & \underline{\C}(\bm{u})\ {\rm is}\ e\mbox{-}{\rm Galois\  dual}\mbox{-}{\rm containing}\\
\Leftrightarrow\ & \underline{\C}(\bm{u})^{\perp_e}\subseteq \underline{\C}(\bm{u})
\Leftrightarrow \sigma^{m-e}(\underline{\C}(\bm{u})^{\bot_E})\subseteq \underline{\C}(\bm{u}) \\
\Leftrightarrow\ & \underline{\C}(\bm{u})^{\bot_E}\subseteq \sigma^e( \underline{\C}(\bm{u}))
\Leftrightarrow \underline{\C}^{\bot_E}[\bm{u}]\subseteq \sigma^e( (\underline{\C}(\bm{u})^{\bot_E})^{\bot_E})
 \\
\Leftrightarrow\ & \underline{\C}^{\bot_E}[\bm{u}]\subseteq (\underline{\C}^{\bot_E}[\bm{u}])^{\bot_{m-e}}\\
\Leftrightarrow\ &\underline{\C}^{\bot_E}[\bm{u}]\ {\rm is}\ (m-e)\mbox{-}{\rm Galois\  self}\mbox{-}{\rm orthogonal}\\
\Leftrightarrow\ &\underline{\C}^{\bot_E}[\bm{u}]\ {\rm is}\ e\mbox{-}{\rm Galois\  self}\mbox{-}{\rm orthogonal}.\\
\end{split}\]
By Theorem \ref{th galois 1},
$\underline{\C}^{\bot_E}[\bm{u}]$ is $e$-Galois self-orthogonal if and only if
$\C^{\bot_E}$ is $e$-Galois self-orthogonal, $\bm{u}\in (\C^{\bot_E})^{\bot_{e}}$,
$\sigma^{m-2e}(\bm{u})\in (\C^{\bot_E})^{\bot_{e}}$
and $s(\bm{u}^{p^e+1})=(-1)^{p^e}$.
Note that
 \[\begin{split}  & \C^{\bot_E}\ {\rm is}\ e\mbox{-}{\rm Galois\  self}\mbox{-}{\rm orthogonal}\\
\Leftrightarrow\ & \C^{\bot_E}\ {\rm is}\ (m-e)\mbox{-}{\rm Galois\  self}\mbox{-}{\rm orthogonal}\\
\Leftrightarrow\ & \C^{\bot_E}\subseteq (\C^{\bot_E})^{\bot_{m-e}}
\Leftrightarrow \C^{\bot_{e}}\subseteq \C,\\
\end{split}\]
then the rest proof can be obtained from $(\C^{\bot_E})^{\bot_{e}}=\sigma^{m-e}(\C)$. This completes the proof.
\end{proof}

\section{The existence of Galois self-dual GRS and TGRS codes}\label{sec3}

\subsection{Galois self-dual GRS and EGRS codes}

In this section, we give the main results on the existence of $e$-Galois self-dual GRS and EGRS codes.
Now, we recall the definitions of GRS and EGRS codes.

\begin{definition}
(\cite{RefJ (2003) W. Fund})
Let $\aaa=(\as_1,\as_2,\dots,\as_n)\in \F_q^n$ with $\as_i\neq \as_j\ (i\neq j)$ and $\bm{v}=(v_1,v_2,\dots,v_n)\in (\F_q^*)^n$.
For an integer $k$ satisfying $1\leq k\leq n$, the GRS code associated to $\aaa$ and $\bm{v}$ is defined as
$$GRS_{k}(\aaa,\bm{v})= \Big\{(v_{1}f(\as_{1}),v_{2}f(\as_{2}),\dots,v_{n}f(\as_{n})):\ f(x)\in \F_{q}[x],\ \deg(f(x))\leq k-1 \Big\}.$$
Moreover, the $k$-dimensional EGRS codes of length $n+1$ associated to $\aaa$ and $\bm{v}$ is defined as
$$GRS_{k}(\aaa,\bm{v},\infty)=\Big \{(v_{1}f(\as_{1}),v_{2}f(\as_{2}),\dots,v_{n}f(\as_{n}),f_{k-1}):\ f(x)\in \F_{q}[x],\ \deg(f(x))\leq k-1 \Big\},$$
where $f_{k-1}$ is the coefficient of $x^{k-1}$ in $f(x)$.
\end{definition}

The elements $\as_1, \as_2,\dots,\as_n$ are called the \emph{code locators} (also known as \emph{evaluation sets}) of
$GRS_k(\aaa,\bm{v})$, and the elements $v_1,v_2,\dots, v_n$ are called the
\emph{column multipliers} of $GRS_k(\aaa,\bm{v})$.
A generator matrix of $GRS_{k}(\aaa,\bm{v})$ is given by
$$G_{k}(\aaa,\bm{v})=(\aaa^0\star \bm{v},\aaa^1\star \bm{v},\dots,\aaa^{k-1}\star \bm{v})^\mathcal{T}.$$
Moreover, the code $GRS_k(\aaa,\bm{v},\infty)$ has a generator matrix
$$G_{k}(\aaa,\bm{v},\infty)=[(\aaa^0\star \bm{v},\aaa^1\star \bm{v},\dots,\aaa^{k-1}\star \bm{v})^\mathcal{T}|\infty_{k-1}^T],$$
where $\infty_{k-1}=(0,\dots,0,1)\in \F_q^k$.
It is well known that GRS codes, EGRS codes and their dual codes are MDS codes.
When $n\leq q$, we have that GRS codes are equivalent to EGRS codes.

\begin{lemma}\label{lem GRS=EGRS}
(\cite{RefJ (1996) Pellikaan})
If $n\leq q$, then a linear code with length $n$ is GRS if and only if it is EGRS.
\end{lemma}

Furthermore, we know that the $e$-Galois dual code of a GRS code is still a GRS code.

\begin{lemma}\label{lem seita GRS}
(\cite{RefJ (2024) Qian galois})
Let $\aaa=(\as_1,\as_2,\dots,\as_n)\in \F_q^n$ with $\as_i\neq \as_j\ (i\neq j)$ and $\bm{v}=(v_1,v_2,\dots,v_n)\in (\F_q^*)^n$,
then
$$GRS_k(\aaa,\bm{v})^{\bot_e}=GRS_{n-k}(\sigma^{m-e}(\aaa),\sigma^{m-e}(\bm{v}^{-1}\star L_{\aaa}^{-1})).$$
\end{lemma}

The following lemma shows that EGRS codes are dual extended codes of GRS codes.

\begin{lemma}\label{lem GRS}
Let $\aaa=(\as_1,\as_2,\dots,\as_n)\in \F_q^n$ with $\as_i\neq \as_j\ (i\neq j)$ and $\bm{v}=(v_1,v_2,\dots,v_n)\in (\F_q^*)^n$,
then
$$GRS_k(\aaa,\bm{v},\infty)=\underline{GRS_{k-1}(\aaa,\bm{v})}[-\aaa^{k-1}\star \bm{v}].$$
\end{lemma}

\begin{proof}
Note that codes $GRS_k(\aaa,\bm{v},\infty)$ and $GRS_{k-1}(\aaa,\bm{v})$ have generator matrices
$$G_{k}(\aaa,\bm{v},\infty)=[(\aaa^0\star \bm{v},\aaa^1\star \bm{v},\dots,\aaa^{k-1}\star \bm{v})^\mathcal{T}|\infty_{k-1}^T],$$
where $\infty_{k-1}=(0,\dots,0,1)\in \F_q^k$,
and
$$G_{k-1}(\aaa,\bm{v})=(\aaa^0\star \bm{v},\aaa^1\star \bm{v},\dots,\aaa^{k-2}\star \bm{v})^\mathcal{T},$$
respectively.
From the definition of dual extended code, we can get the desired results.
\end{proof}

Next we give a sufficient and necessary condition for an EGRS code to be $e$-Galois self-dual.

\begin{theorem}\label{th galois EGRS}
Let $n$ be odd, $\aaa=(\as_1,\as_2,\dots,\as_n)\in \F_q^n$ with $\as_i\neq \as_j\ (i\neq j)$ and $\bm{v}=(v_1,v_2,\dots,v_n)\in (\F_q^*)^n$.
Then $GRS_{\frac{n+1}{2}}(\aaa,\bm{v},\infty)$ is $e$-Galois self-dual if and only if
$$GRS_{\frac{n+1}{2}}(\aaa,\bm{v})=GRS_{\frac{n-1}{2}}(\aaa,\bm{v})^{\bot_e},$$
$\sigma^{m-2e}(\aaa^{\frac{n-1}{2}}\star \bm{v})\in GRS_{\frac{n-1}{2}}(\aaa,\bm{v})^{\bot_e}$
and $s((\aaa^{\frac{n-1}{2}}\star \bm{v})^{p^e+1})=-1$.
\end{theorem}

\begin{proof}
By Theorem \ref{th galois 1} and Lemma \ref{lem GRS}, it follows that $GRS_{\frac{n+1}{2}}(\aaa,\bm{v},\infty)$ is $e$-Galois self-dual if and only if
$GRS_{\frac{n-1}{2}}(\aaa,\bm{v})\subseteq GRS_{\frac{n-1}{2}}(\aaa,\bm{v})^{\bot_e}$,
$$\aaa^{\frac{n-1}{2}}\star \bm{v},\ \sigma^{m-2e}(\aaa^{\frac{n-1}{2}}\star \bm{v})\in GRS_{\frac{n-1}{2}}(\aaa,\bm{v})^{\bot_e}$$ and $s((\aaa^{\frac{n-1}{2}}\star \bm{v})^{p^e+1})=-1$.
Combining
 $$GRS_{\frac{n-1}{2}}(\aaa,\bm{v})\subseteq GRS_{\frac{n-1}{2}}(\aaa,\bm{v})^{\bot_e}\quad {\rm and}\quad
\aaa^{\frac{n-1}{2}}\star \bm{v}\in GRS_{\frac{n-1}{2}}(\aaa,\bm{v})^{\bot_e},$$
we have that
$GRS_{\frac{n+1}{2}}(\aaa,\bm{v})\subseteq GRS_{\frac{n-1}{2}}(\aaa,\bm{v})^{\bot_e}$.
Note that
 $$\dim(GRS_{\frac{n+1}{2}}(\aaa,\bm{v}))=\dim(GRS_{\frac{n-1}{2}}(\aaa,\bm{v})^{\bot_e})=\frac{n+1}{2},$$
then $GRS_{\frac{n+1}{2}}(\aaa,\bm{v})=GRS_{\frac{n-1}{2}}(\aaa,\bm{v})^{\bot_e}$.  This completes the proof.
\end{proof}

By Lemma \ref{lem seita GRS} and Theorem \ref{th galois EGRS}, we can directly obtain the following corollary.

\begin{corollary}
Let $n$ be odd, $\aaa=(\as_1,\as_2,\dots,\as_n)\in \F_q^n$ with $\as_i\neq \as_j\ (i\neq j)$ and $\bm{v}=(v_1,v_2,\dots,v_n)\in (\F_q^*)^n$.
Then $GRS_{\frac{n+1}{2}}(\aaa,\bm{v},\infty)$ is $e$-Galois self-dual if and only if
$$GRS_{\frac{n+1}{2}}(\aaa,\bm{v})=GRS_{\frac{n+1}{2}}(\sigma^{m-e}(\aaa),\sigma^{m-e}(\bm{v}^{-1}\star L_{\aaa}^{-1})),$$
$\aaa^{\frac{n-1}{2}}\star \bm{v}\in GRS_{\frac{n+1}{2}}(\sigma^{e}(\aaa),\sigma^{e}(\bm{v}^{-1}\star L_{\aaa}^{-1}))$
and $s((\aaa^{\frac{n-1}{2}}\star \bm{v})^{p^e+1})=-1$.
\end{corollary}

The next theorem shows that some lengths of $e$-Galois self-dual (extended) GRS codes do not exist.

\begin{theorem}\label{th GRS not exist}
Let $n$ be odd with $n\geq 3$. If $1\leq e\leq m-1$ and $n>\min\{p^e,p^{m-e}\}$, then there is no $e$-Galois self-dual (extended) GRS codes of length $n+1$.
\end{theorem}

\begin{proof}
Note that the length of GRS codes is at most $q$, then by Lemma \ref{lem GRS=EGRS}, we only need to prove the case of EGRS codes.
Let $\aaa=(\as_1,\as_2,\dots,\as_n)\in \F_q^n$ with $\as_i\neq \as_j\ (i\neq j)$ and $\bm{v}=(v_1,v_2,\dots,v_n)\in (\F_q^*)^n$.
Assume that there exists an $e$-Galois self-dual EGRS code $GRS_{\frac{n+1}{2}}(\aaa,\bm{v},\infty)$.
Now based on the length $n$, we divide the proof into the following two parts.

$\bullet$ $\textbf{Case 1:}$ $n\geq 2p^{m-e}+1$.

By Lemma \ref{lem galois OA}, we know that $GRS_{\frac{n+1}{2}}(\aaa,\bm{v},\infty)$ is $(m-e)$-Galois self-dual.
Note that the vectors $(\bm{a}^i\star \bm{v},0),\ 0\leq i\leq \frac{n-3}{2}$ and $(\bm{a}^{\frac{n-1}{2}}\star \bm{v},1)$ form a basis of $GRS_{\frac{n+1}{2}}(\bm{a},\bm{v},\infty)$.
By definition, it follows
$$s(\aaa^{i+p^{m-e}j}\star \bm{v}^{p^{m-e}+1})=0,\ {\rm for}\ 0\leq i,\ j\leq \frac{n-1}{2},\  {\rm except}\  i=j=\frac{n-1}{2}.$$
Since  $n\geq 2p^{m-e}+1$, then $\frac{n-1}{2}\geq p^{m-e}$,
 which implies that
 $$s(\aaa^{i}\star \bm{v}^{p^{m-e}+1})=0,\ {\rm for}\ 0\leq i\leq \frac{n-1}{2}(p^{m-e}+1)-1.$$
Since
$n\geq 3$ and $1\leq e\leq m-1$,
then $n\geq 1+\lceil \frac{2}{p^{m-e}-1}\rceil$,
it follows that $n\leq \frac{n-1}{2}(p^{m-e}+1)$.
Hence, we have
 $$s(\aaa^{i}\star \bm{v}^{p^{m-e}+1})=0,\ {\rm for}\ 0\leq i\leq n-1.$$
Note that the set of equations  $(\aaa^0,\aaa^1,\dots,\aaa^{n-1})^{\mathcal{T}}\bm{x}^T=\bm{0}^T$ has only zero solutions.
This contradicts $\bm{v}\in (\F_q^*)^n$, so we have $n< 2p^{m-e}+1$.

$\bullet$ $\textbf{Case 2:}$ $n< 2p^{m-e}+1$.

By Theorem \ref{th galois EGRS},
we have
$GRS_{\frac{n+1}{2}}(\aaa,\bm{v})=GRS_{\frac{n-1}{2}}(\aaa,\bm{v})^{\bot_e}$.
Note that
 \[\begin{split}  & GRS_{\frac{n+1}{2}}(\aaa,\bm{v})=GRS_{\frac{n-1}{2}}(\aaa,\bm{v})^{\bot_e}\\
\Leftrightarrow\ & GRS_{\frac{n+1}{2}}(\aaa,\bm{v})=\sigma^{m-e}(GRS_{\frac{n-1}{2}}(\aaa,\bm{v})^{\bot_E})\\
\Leftrightarrow\ & \sigma^e(GRS_{\frac{n+1}{2}}(\aaa,\bm{v}))=GRS_{\frac{n-1}{2}}(\aaa,\bm{v})^{\bot_E}\\
\Leftrightarrow\ & GRS_{\frac{n+1}{2}}(\sigma^e(\aaa),\sigma^e(\bm{v}))=GRS_{\frac{n-1}{2}}(\aaa,\bm{v})^{\bot_E},\\
\end{split}\]
then we have $GRS_{\frac{n+1}{2}}(\sigma^e(\aaa),\sigma^e(\bm{v}))=GRS_{\frac{n-1}{2}}(\aaa,\bm{v})^{\bot_E}$.
Next, we will prove that $\aaa\star \sigma^e(\bm{v})\in GRS_{\frac{n-1}{2}}(\aaa,\bm{v})^{\bot_E}$.
It suffices to check that
$$s(\aaa^{i}\star \sigma^e(\bm{v})\star \bm{v})=0,\ {\rm for}\ 1\leq i\leq \frac{n-1}{2}.$$
Note that $\sigma^e(\bm{v})\in GRS_{\frac{n-1}{2}}(\aaa,\bm{v})^{\bot_E}$,
it follows that
$$s(\aaa^{i}\star \sigma^e(\bm{v})\star \bm{v})=0,\ {\rm for}\ 1\leq i\leq \frac{n-3}{2}.$$
Therefore, we only need to prove that $s(\aaa^{\frac{n-1}{2}}\star \sigma^e(\bm{v})\star \bm{v})=0$.
By Theorem \ref{th galois EGRS}, we have that $\sigma^{m-2e}(\aaa^{\frac{n-1}{2}}\star \bm{v})\in GRS_{\frac{n-1}{2}}(\aaa,\bm{v})^{\bot_e}$,
it follows that
$$G_{\frac{n-1}{2}}(\aaa,\bm{v})\sigma^{m-e}(\aaa^{\frac{n-1}{2}}\star \bm{v})^T=\bm{0}^T
\Leftrightarrow G_{\frac{n-1}{2}}(\sigma^e(\aaa),\sigma^e(\bm{v}))(\aaa^{\frac{n-1}{2}}\star \bm{v})^T=\bm{0}^T.$$
From this we have $s(\aaa^{\frac{n-1}{2}}\star \sigma^e(\bm{v})\star \bm{v})=0$.
Hence,
$$\aaa\star \sigma^e(\bm{v})\in GRS_{\frac{n-1}{2}}(\aaa,\bm{v})^{\bot_E}=GRS_{\frac{n+1}{2}}(\sigma^e(\aaa),\sigma^e(\bm{v})).$$
Then there exist $l_0,l_1,\dots,l_{\frac{n-1}{2}}$  which are not all zero such that
$$
\aaa\star \sigma^e(\bm{v})=l_0\sigma^e(\bm{v})+l_1\sigma^e(\aaa)\star \sigma^e(\bm{v})+\dots+l_{\frac{n-1}{2}}
\sigma^e(\aaa^{\frac{n-1}{2}})\star \sigma^e(\bm{v}).
$$
Both sides of the equation are raised to the $p^{m-e}$ power at the same time, and we get
$$
\sigma^{m-e}(\aaa)\star \bm{v}=\sigma^{m-e}(l_0)\bm{v}+\sigma^{m-e}(l_1)\aaa\star \bm{v}+\dots+\sigma^{m-e}(l_{\frac{n-1}{2}})
\aaa^{\frac{n-1}{2}}\star \bm{v}.
$$
It follows that
$$
\sigma^{m-e}(\aaa)=\sigma^{m-e}(l_0)\bm{1}+\sigma^{m-e}(l_1)\aaa+\dots+\sigma^{m-e}(l_{\frac{n-1}{2}})
\aaa^{\frac{n-1}{2}}.
$$
This implies that the polynomial
$$g(x)=x^{p^{m-e}}-\sigma^{m-e}(l_{\frac{n-1}{2}})
x^{\frac{n-1}{2}}-\sigma^{m-e}(l_{\frac{n-3}{2}})x^{\frac{n-3}{2}}-\dots-\sigma^{m-e}(l_0)$$
has $n$ distinct roots $\as_1, \as_2,\dots, \as_{n}$.
Since $n< 2p^{m-e}+1$, then $p^{m-e}> \frac{n-1}{2}$, which implies that $\deg(g(x))= p^{m-e}$
and $g(x)$ is not a zero polynomial.
From the fact that $\as_1, \as_2, \dots, \as_{n}$
are distinct roots of the polynomial $g(x)$, we have $n\leq p^{m-e}$.

Combining $\textbf{Case 1}$ and $\textbf{Case 2}$, it is not difficult to deduce that $n\leq p^{m-e}$.
By Lemma \ref{lem galois OA},
we have that $GRS_{\frac{n+1}{2}}(\aaa,\bm{v},\infty)$ is $e$-Galois self-dual if and only if $GRS_{\frac{n+1}{2}}(\aaa,\bm{v},\infty)$ is $(m-e)$-Galois self-dual.
Similarly, we have $n\leq p^{e}$.
It follows that $n\leq \min\{p^e,p^{m-e}\}$.
This completes the proof.
\end{proof}

\begin{remark}
By the result of Theorem \ref{th GRS not exist}, the studies listed below are valuable and interesting.
\begin{itemize}
\item
By the proof of Theorem \ref{th GRS not exist}, we know that all elements in the evaluation sets of $e$-Galois self-dual EGRS codes  are roots of the polynomial $g(x)$
of the form
$$g(x)=x^{p^{e}}-l_{\frac{n-1}{2}} x^{\frac{n-1}{2}}-l_{\frac{n-3}{2}}x^{\frac{n-3}{2}}-\dots-l_0,$$ where
$l_0,l_1,\dots, l_{\frac{n-1}{2}}$ are not all zero.
Therefore, if the number of roots of the polynomial $g(x)$ over $\F_{q}$ can be determined,
the non-existence of the $e$-Galois self-dual (extended) GRS code can also be determined.

\item
Furthermore, is it possible to determine the polynomial $g(x)$ in the form of
$$g(x)=x^{p^{e}}-l_1x-l_0,$$ where $l_1\neq 0$?
If so, according to the results in \cite{RefJ (2004) x^p-ax-b},
it follows that there does not exist an $e$-Galois self-dual (extended) GRS code of length $n> p^{\gcd(e,m)}+1$, where $1\leq e\leq m-1$.
\end{itemize}
\end{remark}

In \cite[Corollary 4]{RefJ (2024) wan so}, the authors constructed some classes of $e$-Galois self-dual (extended) GRS codes. We combine this with Theorem \ref{th GRS not exist} to obtain the following corollary, which shows that for some cases, Theorem \ref{th GRS not exist} gives the maximum length of $e$-Galois self-dual (extended) GRS codes.

\begin{corollary}\label{cor 33}
Let $q=p^m$ be a prime power and $e$ be an integer
with $e\mid m$.
Let $n$ be even with $n\geq 4$.
Then the following two assertions hold.

\begin{itemize}
\item[(1)]
Suppose $\frac{m}{e}$ is odd and $p=2$, then
there exists a $q$-ary $e$-Galois self-dual (extended) GRS code
of length $n$ if and only if $n\leq p^e$.
\item[(2)]
Suppose $\frac{m}{e}$ is even, then
there exists a $q$-ary $e$-Galois self-dual (extended) GRS code
of length $n$ if and only if $n\leq p^e+1$.
\end{itemize}
\end{corollary}

\begin{remark}
By Lemma \ref{lem galois OA}, similarly to Corollary \ref{cor 33},
we can directly obtain the case $(m-e)\mid m$, which we omit.
Moreover, Corollary \ref{cor 33} implies that for some cases we can solve the problem of the existence of $e$-Galois self-dual (extended) GRS codes.
\end{remark}

\subsection{Galois self-dual TGRS and ETGRS codes}

In this section, we give the main results on the existence of $e$-Galois self-dual TGRS and ETGRS codes.
The definition of the twisted polynomials linear space $\mathcal{V}_{k,t,h,\eta,q}$ is given in the following.

\begin{definition}
(\cite{RefJ (2017) Bee TGRS})
Let $\eta\in \F_q^*$ and $t$, $h$, $k$ be integers, with $0\leq h<k\leq q$.
Then the set of $(k,t,h,\eta)$-twisted polynomials is defined as
$$\mathcal{V}_{k,t,h,\eta,q}=\Big \{f(x)=\sum_{i=0}^{k-1}f_ix^i+\eta f_h x^{k-1+t}:\ f_i\in \F_q\ (0\leq i\leq k-1) \Big \},$$
which is a $k$-dimensional $\F_q$-linear subspace.
$h$ and $t$ are the hook and the twist, respectively.
\end{definition}

From the twisted polynomials linear space $\mathcal{V}_{k,t,h,\eta,q}$, the definitions of the TGRS code and the ETGRS code are given in the following, respectively.

\begin{definition}
(\cite{RefJ (2017) Bee TGRS})
For any integers $t$, $h$, $k$, $n$ with $0\leq h\leq k-1<k-1+t<n\leq q$,
let $\eta \in \F_q^*$, $\aaa=(\as_1,\as_2,\dots,\as_n)\in \F_q^n$ with $\as_i\neq \as_j$ $(i\neq j)$ and $\bm{v}=(v_1,v_2,\dots,v_n)\in (\F_q^*)^n$.
Then the TGRS code is defined as
$$TGRS_{k,t,h,\eta}(\aaa,\bm{v})=\Big \{(v_{1}f(\as_{1}),v_{2}f(\as_{2}),\dots,v_{n}f(\as_{n})):\ f(x)\in \mathcal{V}_{k,t,h,\eta,q} \Big \}.$$
Moreover, for $h>0$, the ETGRS code is defined as
$$TGRS_{k,t,h,\eta}(\aaa,\bm{v},\infty)= \Big \{(v_{1}f(\as_{1}),v_{2}f(\as_{2}),\dots,v_{n}f(\as_{n}),f_h):\ f(x)\in \mathcal{V}_{k,t,h,\eta,q} \Big\},$$
where $f_h$ is the coefficient of $x^h$ in $f(x)$.
\end{definition}

By the definition of $TGRS$ and $ETGRS$ code,
a generator matrix of $TGRS_{k,t,h,\eta}(\aaa,\bm{v})$ is given by
$$G_{k,t,h,\eta}(\aaa,\bm{v})=(\aaa^0\star \bm{v},\dots,\aaa^{h-1}\star \bm{v},(\aaa^h+\eta \aaa^{k-1+t})\star \bm{v}, \aaa^{h+1}\star \bm{v}, \dots, \aaa^{k-1}\star \bm{v})^\mathcal{T}.$$
Moreover, $TGRS_{k,t,h,\eta}(\aaa,\bm{v},\infty)$ has the following generator matrix
$$G_{k,t,h,\eta}(\aaa,\bm{v},\infty)=[(\aaa^0\star \bm{v},\dots,\aaa^{h-1}\star \bm{v},(\aaa^h+\eta \aaa^{k-1+t})\star \bm{v}, \aaa^{h+1}\star \bm{v}, \dots, \aaa^{k-1}\star \bm{v})^\mathcal{T}| \infty_{h}^T],$$
where $\infty_h=(\underbrace{0,\dots,0}_{h\ times},1,0,\dots,0)\in \F_q^k$.

In this subsection, we only discuss TGRS codes with $h=0$ and ETGRS codes with $h=k-1$.

\subsubsection{Galois self-dual TGRS code with $h=0$ and $0\in A_{\aaa}$}

The following lemma shows that TGRS codes are dual extended codes of GRS codes if $h=0$ and $0\in A_{\aaa}$.
If $\bm{v}_1, \bm{v}_2\in (\F_q^*)^n$ and $\aaa_1, \aaa_2\in \F_q^n$ with $A_{\aaa_1}=A_{\aaa_2}$,
then $TGRS_{k,t,0,\eta}(\aaa_1,\bm{v}_1)$ and $TGRS_{k,t,0,\eta}(\aaa_2,\bm{v}_2)$ are equivalent.
Thus, for $0\in A_{\aaa}$, we always set $\as_n=0$ without loss of generality.

\begin{lemma}\label{lem *TGRS}
Let $\aaa=(\as_1,\as_2,\dots,\as_n=0)\in \F_q^n$ with $\as_i\neq \as_j\ (i\neq j)$, and $\bm{v}=(v_1,v_2,\dots,v_n)\in (\F_q^*)^n$.
Let $\tilde{\aaa}=(\as_1,\as_2,\dots,\as_{n-1})\in (\F_q^*)^{n-1}$ and $\tilde{\bm{v}}=(v_1,v_2,\dots,v_{n-1})\in (\F_q^*)^{n-1}$.
Then
$$TGRS_{k,t,0,\eta}(\aaa,\bm{v})=\underline{GRS_{k-1}(\tilde{\aaa},\tilde{\aaa}\star \tilde{\bm{v}})}[-v_{n}^{-1}(\tilde{\aaa}^0+\eta\tilde{\aaa}^{k-1+t})\star \tilde{\bm{v}}].$$
\end{lemma}

\begin{proof}
By definition, the code $TGRS_{k,t,0,\eta}(\aaa,\bm{v})$ has a generator matrix
\[\begin{split}
G_{k,t,0,\eta}(\aaa,\bm{v})&=
((\aaa^0+\eta\aaa^{k-1+t})\star \bm{v},\aaa\star \bm{v}, \dots, \aaa^{k-1}\star \bm{v})^\mathcal{T}\\
& =
[(\tilde{\aaa}^0+\eta\tilde{\aaa}^{k-1+t})\star \tilde{\bm{v}},\tilde{\aaa}\star \tilde{\bm{v}},\dots,\tilde{\aaa}^{k-1}\star \tilde{\bm{v}})^\mathcal{T} |v_{n}\infty_0^T],\\
\end{split}\]
where $\infty_0=(1,0,\dots,0)\in \F_q^k$.
Note that $GRS_{k-1}(\tilde{\aaa},\tilde{\aaa}\star \tilde{\bm{v}})$ has a generator matrix
$$G_{k-1}(\tilde{\aaa},\tilde{\aaa}\star \tilde{\bm{v}})=(\tilde{\aaa}\star \tilde{\bm{v}},\tilde{\aaa}^2\star \tilde{\bm{v}},\dots,\tilde{\aaa}^{k-1}\star \tilde{\bm{v}})^{\mathcal{T}}.$$
From the definition of dual extended code, we can get the desired results.
\end{proof}

Next we give a sufficient and  necessary condition for an TGRS code to be $e$-Galois self-dual when
$h=0$ and $0\in A_{\aaa}$.

\begin{theorem}\label{th galois *TGRS}
Let $n$ be even, $\aaa=(\as_1,\as_2,\dots,\as_n=0)\in \F_q^n$ with $\as_i\neq \as_j\ (i\neq j)$ and $\bm{v}=(v_1,v_2,\dots,v_n)\in (\F_q^*)^n$.
Let $\tilde{\aaa}=(\as_1,\as_2,\dots,\as_{n-1})\in (\F_q^*)^{n-1}$ and $\tilde{\bm{v}}=(v_1,v_2,\dots,v_{n-1})\in (\F_q^*)^{n-1}$.
Then $TGRS_{\frac{n}{2},t,0,\eta}(\aaa,\bm{v})$ is $e$-Galois self-dual if and only if
$$TGRS_{\frac{n}{2},t,0,\eta}(\tilde{\aaa},\tilde{\bm{v}})=GRS_{\frac{n}{2}-1}(\tilde{\aaa},\tilde{\aaa}\star \tilde{\bm{v}})^{\bot_e},$$
$\sigma^{m-2e}((\tilde{\aaa}^0+\eta\tilde{\aaa}^{\frac{n}{2}-1+t})\star\tilde{\bm{v}})\in GRS_{\frac{n}{2}-1}(\tilde{\aaa},\tilde{\aaa}\star \tilde{\bm{v}})^{\bot_e}$
and $s(((\tilde{\aaa}^0+\eta\tilde{\aaa}^{\frac{n}{2}-1+t})\star \tilde{\bm{v}})^{p^e+1})=-v_{n}^{p^e+1}$.
\end{theorem}

\begin{proof}
By Theorem \ref{th galois 1} and Lemma \ref{lem *TGRS}, we have that $TGRS_{\frac{n}{2},t,0,\eta}(\aaa,\bm{v})$ is $e$-Galois self-dual if and only if
$GRS_{\frac{n}{2}-1}(\tilde{\aaa},\tilde{\aaa}\star \tilde{\bm{v}})\subseteq GRS_{\frac{n}{2}-1}(\tilde{\aaa},\tilde{\aaa}\star \tilde{\bm{v}})^{\bot_e}$,
$$(\tilde{\aaa}^0+\eta\tilde{\aaa}^{\frac{n}{2}-1+t})\star \tilde{\bm{v}},\ \sigma^{m-2e}((\tilde{\aaa}^0+\eta\tilde{\aaa}^{\frac{n}{2}-1+t})\star \tilde{\bm{v}})\in GRS_{\frac{n}{2}-1}(\tilde{\aaa},\tilde{\aaa}\star \tilde{\bm{v}})^{\bot_e}$$
 and $s(((\tilde{\aaa}^0+\eta\tilde{\aaa}^{\frac{n}{2}-1+t})\star \tilde{\bm{v}})^{p^e+1})=-v_{n}^{p^e+1}$.
Combining
$$GRS_{\frac{n}{2}-1}(\tilde{\aaa},\tilde{\aaa}\star \tilde{\bm{v}})\subseteq GRS_{\frac{n}{2}-1}(\tilde{\aaa},\tilde{\aaa}\star \tilde{\bm{v}})^{\bot_e}\
{\rm and} \ (\tilde{\aaa}^0+\eta\tilde{\aaa}^{\frac{n}{2}-1+t})\star \tilde{\bm{v}}\in GRS_{\frac{n}{2}-1}(\tilde{\aaa},\tilde{\aaa}\star \tilde{\bm{v}})^{\bot_e},
$$
we have that
$TGRS_{\frac{n}{2},t,0,\eta}(\tilde{\aaa},\tilde{\bm{v}})\subseteq GRS_{\frac{n}{2}-1}(\tilde{\aaa},\tilde{\aaa}\star \tilde{\bm{v}})^{\bot_e}$.
Since $$\dim(TGRS_{\frac{n}{2},t,0,\eta}(\tilde{\aaa},\tilde{\bm{v}}))=\dim(GRS_{\frac{n}{2}-1}(\tilde{\aaa},\tilde{\aaa}\star \tilde{\bm{v}})^{\bot_e})=\frac{n}{2},$$
then $TGRS_{\frac{n}{2},t,0,\eta}(\tilde{\aaa},\tilde{\bm{v}})=GRS_{\frac{n}{2}-1}(\tilde{\aaa},\tilde{\aaa}\star \tilde{\bm{v}})^{\bot_e}$.
The following proof is similar to Theorem \ref{th galois EGRS} and we omit the details.
 This completes the proof.
\end{proof}

Similarly, by Lemma \ref{lem seita GRS} and Theorem \ref{th galois *TGRS}, we can directly obtain the following corollary.

\begin{corollary}\label{cor h=0 TGRS}
Let $n$ be even, $\aaa=(\as_1,\as_2,\dots,\as_n=0)\in \F_q^n$ with $\as_i\neq \as_j\ (i\neq j)$ and $\bm{v}=(v_1,v_2,\dots,v_n)\in (\F_q^*)^n$.
Let $\tilde{\aaa}=(\as_1,\as_2,\dots,\as_{n-1})\in (\F_q^*)^{n-1}$ and $\tilde{\bm{v}}=(v_1,v_2,\dots,v_{n-1})\in (\F_q^*)^{n-1}$.
Then $TGRS_{\frac{n}{2},t,0,\eta}(\aaa,\bm{v})$ is $e$-Galois self-dual if and only if
$$TGRS_{\frac{n}{2},t,0,\eta}(\tilde{\aaa},\tilde{\bm{v}})= GRS_{\frac{n}{2}}(\sigma^{m-e}(\tilde{\aaa}),\sigma^{m-e}(\tilde{\aaa}^{-1}\star \tilde{\bm{v}}^{-1}\star L_{\tilde{\aaa}}^{-1})),$$
$(\tilde{\aaa}^0+\eta\tilde{\aaa}^{\frac{n}{2}-1+t})\star\tilde{\bm{v}}\in GRS_{\frac{n}{2}}(\sigma^{e}(\tilde{\aaa}),\sigma^{e}(\tilde{\aaa}^{-1}\star \tilde{\bm{v}}^{-1}\star L_{\tilde{\aaa}}^{-1}))$
and $s(((\tilde{\aaa}^0+\eta\tilde{\aaa}^{\frac{n}{2}-1+t})\star \tilde{\bm{v}})^{p^e+1})=-v_{n}^{p^e+1}$.
Hence, if
$TGRS_{\frac{n}{2},t,0,\eta}(\tilde{\aaa},\tilde{\bm{v}})$
is non-GRS, then $TGRS_{\frac{n}{2},t,0,\eta}(\aaa,\bm{v})$ is not $e$-Galois self-dual.
\end{corollary}

\begin{remark}
The determination of the non-GRS properties of TGRS codes is an interesting problem.
This problem has received significant attention in the literature (see, e.g., \cite{RefJ (2017) Bee TGRS}, \cite{RefJ (2022) Bee TGRS}, \cite{RefJ (2021) H. Liu dcc}, \cite{RefJ (2022) Zhu non-GRS}, \cite{RefJ (2024) Zhu*}).
 The new results on the non-GRS properties of TGRS codes can be used to prove the existence of the corresponding $e$-Galois self-dual TGRS codes by applying Corollary \ref{cor h=0 TGRS}.
\end{remark}

\begin{lemma}\label{lem h=0 not TGRS}
(\cite{RefJ (2022) Zhu non-GRS} and \cite{RefJ (2024) Zhu*})
Suppose that $\tilde{\aaa}$ and $\tilde{\bm{v}}$ are defined as above.
Let $n$ be even and $t$ be integer, with $n\geq 8$, $1\leq t\leq \frac{n-6}{2}$ and $t\neq 2$, then $TGRS_{\frac{n}{2},t,0,\eta}(\tilde{\aaa},\tilde{\bm{v}})$
is non-GRS.
\end{lemma}

\begin{proof}
For the case $t=1$, the required results have been documented in \cite[Theorem 3.2]{RefJ (2024) Zhu*}.
For the case $3\leq t\leq \frac{n-6}{2}$, the desired result can be obtained from \cite[Theorem 1 (1.5)]{RefJ (2022) Zhu non-GRS}.
\end{proof}

By Lemma \ref{lem h=0 not TGRS} and Corollary \ref{cor h=0 TGRS}, we can directly obtain the following theorem.

\begin{theorem}\label{th TGRS not exist}
Let $n$ be even and $t$ be integer, with $n\geq 8$, $1\leq t\leq \frac{n-6}{2}$ and $t\neq 2$.
If $h=0$ and $0\in A_{\aaa}$, then there is no $e$-Galois self-dual TGRS code of length $n$.
\end{theorem}

\subsubsection{Galois self-dual ETGRS code with $h=k-1$}

The following lemma shows that ETGRS codes are dual extended codes of GRS codes if $h=k-1$.

\begin{lemma}\label{lem ETGRS}
Let $\aaa=(\as_1,\as_2,\dots,\as_n)\in \F_q^n$ with $\as_i\neq \as_j\ (i\neq j)$ and $\bm{v}=(v_1,v_2,\dots,v_n)\in (\F_q^*)^n$,
then
$$TGRS_{k,t,k-1,\eta}(\aaa,\bm{v},\infty)=
\underline{GRS_{k-1}(\aaa, \bm{v})}[-(\aaa^{k-1}+\eta\aaa^{k-1+t})\star \bm{v}].$$
\end{lemma}

\begin{proof}
By definition,
the code $TGRS_{k,t,k-1,\eta}(\aaa,\bm{v},\infty)$ has a generator matrix
$$G_{k,t,k-1,\eta}(\aaa,\bm{v},\infty)=[(\aaa^0\star \bm{v},\dots,\aaa^{k-2}\star \bm{v},(\aaa^{k-1}+\eta \aaa^{k-1+t})\star \bm{v})^\mathcal{T}| \infty_{k-1}^T],$$
where $\infty_{k-1}=(0,\dots,0,1)\in \F_q^k$.
And $GRS_{k-1}(\aaa, \bm{v})$ has a generator matrix
$$G_{k-1}(\aaa,\bm{v})=(\aaa^0 \star \bm{v}, \aaa \star \bm{v},\dots,\aaa^{k-2}\star \bm{v})^{\mathcal{T}}.$$
From the definition of dual extended code, we can get the desired results.
\end{proof}

Similarly, we give a sufficient and necessary condition for an ETGRS code to be $e$-Galois self-dual when
$h=k-1$.

\begin{theorem}\label{th galois ETGRS}
Let $n$ be odd, $\aaa=(\as_1,\as_2,\dots,\as_n)\in \F_q^n$ with $\as_i\neq \as_j\ (i\neq j)$ and $\bm{v}=(v_1,v_2,\dots,v_n)\in (\F_q^*)^n$.
Then $TGRS_{\frac{n+1}{2},t,\frac{n-1}{2},\eta}(\aaa,\bm{v},\infty)$ is $e$-Galois self-dual if and only if
$$TGRS_{\frac{n+1}{2},t,\frac{n-1}{2},\eta}(\aaa,\bm{v})=GRS_{\frac{n-1}{2}}(\aaa,\bm{v})^{\bot_e},$$
$\sigma^{m-2e}((\aaa^{\frac{n-1}{2}}+\eta\aaa^{\frac{n-1}{2}+t})\star\bm{v})\in GRS_{\frac{n-1}{2}}(\aaa, \bm{v})^{\bot_e}$
and $s(((\aaa^{\frac{n-1}{2}}+\eta\aaa^{\frac{n-1}{2}+t})\star \bm{v})^{p^e+1})=-1$.
\end{theorem}

\begin{proof}
By employing  Theorem \ref{th galois 1} and Lemma \ref{lem ETGRS},
the theorem can be proved similarly as Theorem \ref{th galois *TGRS}. We omit the
details. This completes the proof.
\end{proof}

Similarly, by Lemma \ref{lem seita GRS} and Theorem \ref{th galois ETGRS}, we can directly obtain the following corollary.

\begin{corollary}\label{cor h=k-1 TGRS}
Let $n$ be odd, $\aaa=(\as_1,\as_2,\dots,\as_n)\in \F_q^n$ with $\as_i\neq \as_j\ (i\neq j)$ and $\bm{v}=(v_1,v_2,\dots,v_n)\in (\F_q^*)^n$.
Then $TGRS_{\frac{n+1}{2},t,\frac{n-1}{2},\eta}(\aaa,\bm{v},\infty)$ is $e$-Galois self-dual if and only if
$$TGRS_{\frac{n+1}{2},t,\frac{n-1}{2},\eta}(\aaa,\bm{v})=GRS_{\frac{n+1}{2}}(\sigma^{m-e}(\aaa),\sigma^{m-e}( \bm{v}^{-1}\star L_{\aaa}^{-1})),$$
$(\aaa^{\frac{n-1}{2}}+\eta\aaa^{\frac{n-1}{2}+t})\star\bm{v}\in GRS_{\frac{n+1}{2}}(\sigma^{e}(\aaa),\sigma^{e}( \bm{v}^{-1}\star L_{\aaa}^{-1}))$
and $s(((\aaa^{\frac{n-1}{2}}+\eta\aaa^{\frac{n-1}{2}+t})\star \bm{v})^{p^e+1})=-1$.
Hence, if $TGRS_{\frac{n+1}{2},t,\frac{n-1}{2},\eta}(\aaa,\bm{v})$
is non-GRS, then $TGRS_{\frac{n+1}{2},t,\frac{n-1}{2},\eta}(\aaa,\bm{v},\infty)$ is not $e$-Galois self-dual.
\end{corollary}

\begin{remark}
Similarly, the new results on the non-GRS properties of TGRS codes can be used to prove the existence of the corresponding $e$-Galois self-dual ETGRS codes by applying Corollary \ref{cor h=k-1 TGRS}.
\end{remark}

\begin{lemma}\label{lem h=k-1 not TGRS}
(\cite{RefJ (2022) Zhu non-GRS})
Suppose that $\aaa$ and $\bm{v}$ are defined as above.
Let $n$ be odd and $t$ be integer, with $n\geq 11$ and $3\leq t\leq \frac{n-5}{2}$, then $TGRS_{\frac{n+1}{2},t,\frac{n-1}{2},\eta}(\aaa,\bm{v})$
is non-GRS.
\end{lemma}

\begin{proof}
The desired result can be obtained from \cite[Theorem 1 (1.5)]{RefJ (2022) Zhu non-GRS}.
\end{proof}

Similarly, by Lemma \ref{lem h=k-1 not TGRS} and Corollary \ref{cor h=k-1 TGRS}, we can directly obtain the following theorem.

\begin{theorem}\label{th ETGRS not exist}
Let $n$ be odd and $t$ be integer, with $n\geq 11$ and $3\leq t\leq \frac{n-5}{2}$.
 If $h=\frac{n-1}{2}$, then there is no $e$-Galois self-dual ETGRS code of length $n+1$.
\end{theorem}

\section{The construction of Hermitian self-dual $(+)$-TGRS and $(\ast)$-TGRS codes}\label{sec4}

In \cite{RefJ (2017) Bee TGRS}, two explicit constructions of TGRS codes with the case $(t,h)=(1,0)$ and the case $(t,h)=(1,k-1)$ were given.
The codes with $(t,h)=(1,0)$ and $(t,h)=(1,k-1)$ are called $(\ast)$-TGRS codes and $(+)$-TGRS codes, respectively.
In this section, we construct several new classes of Hermitian self-dual (+)-TGRS and ($\ast$)-TGRS codes.
We assume that $\theta$ is a primitive element of $\F_{q^2}$,
that is $\F_{q^2}^*=\langle \theta \rangle$.
For convenience, we write $TGRS_{k,\eta}^{\ast}(\aaa,\bm{v})=TGRS_{k,1,0,\eta}(\aaa,\bm{v})$ and $TGRS_{k,\eta}^{+}(\aaa,\bm{v})=TGRS_{k,1,k-1,\eta}(\aaa,\bm{v})$.
For a vector $\aaa=(\as_1,\as_2,\dots,\as_n)\in \F_{q}^n$, with $\as_i\neq \as_j\ (i\neq j)$,
we always denote $s(\aaa)=\sum_{i=1}^n \as_i$ and $p(\aaa)=\prod_{i=1}^n \as_i$.

\subsection{Hermitian self-dual $(+)$-TGRS codes}

The following lemma shows that $(+)$-TGRS codes is MDS or NMDS.

\begin{lemma}\label{lem +TGRS MDS}
(\cite{RefJ (2017) Bee TGRS})
Let $k\leq n\leq q$,
$\aaa=(\as_1,\as_2,\dots,\as_n)\in \F_{q}^n$, with $\as_i\neq \as_j\ (i\neq j)$, $\bm{v}=(v_1,v_2,\dots,v_n)\in (\F_{q}^*)^n$ and $\eta\in \F_q^*$.
Let $[n]=\{1,2,\dots,n\}$,
then the following two assertions hold.
\begin{itemize}
\item[(1)]
$TGRS_{k,\eta}^{+}(\aaa,\bm{v})$ is MDS if and only if for any $k$-subset $I\subseteq [n]$, it holds
$\eta \sum_{i\in I}\as_i\neq -1$.
\item[(2)]
$TGRS_{k,\eta}^{+}(\aaa,\bm{v})$ is NMDS if and only if
there exists $k$-subset $I\subseteq [n]$ such that $\eta \sum_{i\in I}\as_i=-1$.
\end{itemize}
\end{lemma}

In \cite{RefJ (2023) Guo H-TGRS}, Guo et al. gave a sufficient and  necessary condition for (+)-TGRS codes to be Hermitian self-dual, which we review here.

\begin{lemma}\label{lem +TGRS}
(\cite{RefJ (2023) Guo H-TGRS})
Let $n$ be even, $\aaa=(\as_1,\as_2,\dots,\as_n)\in \F_{q^2}^n$, with $\as_i\neq \as_j\ (i\neq j)$ and $\bm{v}=(v_1,v_2,\dots,v_n)\in (\F_{q^2}^*)^n$,
then $TGRS_{\frac{n}{2},\eta}^{+}(\aaa,\bm{v})$ $(1+s(\aaa)\eta\neq 0)$ over $\F_{q^2}$ is Hermitian self-dual if and only if
there exists a polynomial $g(x)=\sum_{i=0}^{k-2}g_ix^{i}+g_{k-1}(x^{k-1}-\frac{\eta}{1+s(\aaa)\eta}x^k)$,
$g_i\in \F_{q^2}$, $0\leq i\leq k-1$ such that
$$v_i^{q+1}f(\as_i)^q=L_{\aaa}(\as_i)^{-1}g(\as_i),\ 1\leq i\leq n,$$
for all $f(x)\in \mathcal{V}_{\frac{n}{2},1,k-1,\eta,q^2}$.
\end{lemma}

Given a finite set $A$, $\#A$ will denote its cardinality.
Suppose $a\in \F_{q^2}^*$ and $b\in \F_{q^2}$.
Denote a set
 \begin{equation}\label{conn 1}
 S_{a,b}\triangleq \Big \{\alpha\in \F_{q^2}:\ \alpha^q=a\alpha+b \Big\}.
 \end{equation}
It is easy to check that $0\leq \#S_{a,b}\leq q$. In particular, we have $\#S_{a,b}=q$ for some $a$, $b$. For example,
if $a=-1$ and $b\in \F_q$, then $\#S_{-1,b}=q$, and $Tr(\alpha)=b$, where $Tr(x)$ denotes the trace function from $\F_{q^2}$ to $\F_q$.
If $a=1$ and there is some $\alpha\in \F_{q^2}^*$ such that $\alpha^q-\alpha=b$,
then $(\alpha+c)^q=\alpha^q+c^q=\alpha+c+b$, where $c\in \F_q$.
We have that $\{\alpha+c:\ c\in \F_q \}\subseteq S_{1,b}$.
Note that $\#\{\alpha+c:\ c\in \F_q \}=q$, then
$S_{1,b}=\{\alpha+c:\ c\in \F_q \}$.
Then we can choose $a$ and $b$ to make $S_{a,b}$ non-empty.
Let $S_{a,b}$ be defined as in Eq. (\ref{conn 1}), then we obtain the following theorem.

\begin{theorem}\label{th +TGRS}
For given $a\in \F_{q^2}^*$ and $b\in \F_{q^2}$.
Assume that integers $n=2k$ satisfy that there exist $n$ distinct elements $\alpha_1,\alpha_2,\dots,\alpha_n$ such that $\alpha_i\in S_{a,b}$ for $1\leq i\leq n$.
Let $\aaa=(\alpha_1,\alpha_2,\dots,\alpha_n)\in S_{a,b}^n$.
If $\eta\in \F_{q^2}^*$ such that $(a s(\aaa)+kb)\eta^q+a\eta^{q-1}+1=0$,
then there exists a vector $\bm{v}=(v_1,v_2,\dots,v_n)\in (\F_{q^2}^*)^n$ such that $TGRS_{k,\eta}^{+}(\aaa,\bm{v})$ is an $[n,k]_{q^2}$ Hermitian self-dual $(+)$-TGRS code.
\end{theorem}

\begin{proof}
Let $a=\theta^t$ and $\aaa=(\alpha_1,\alpha_2,\dots,\alpha_n)\in S_{a,b}^n$,
then
\[\begin{split}
	L_{\aaa}(\as_i)^{-q} &=\prod_{1\leq j \leq n, j\neq i}(\alpha_i-\alpha_j)^{-q}\\
            &=\prod_{1\leq j \leq n, j\neq i}[(a\alpha_i+b)-(a\alpha_j+b)]^{-1}\\
            &=a^{-(n-1)}\prod_{1\leq j \leq n, j\neq i}(\alpha_i-\alpha_j)^{-1}\\
            &=\theta^{-t(n-1)}L_{\aaa}(\as_i)^{-1}.\\
\end{split}\]
We have $L_{\aaa}(\as_i)^{-(q-1)}=\theta^{-t(n-1)}$,
it follows that $L_{\aaa}(\as_i)^{-1}=\theta^{-\frac{t(n-1)}{q-1}+c(q+1)}$, where $c$ is some integer.
Let $\lambda=\theta^{\frac{t(n-1)}{q-1}}\in \F_{q^2}^*$,
and $v_i=\theta^c$, then we have $v_i^{q+1}=\lambda L_{\aaa}(\as_i)^{-1}$ for all $1\leq i\leq n$.
Set $\bm{v}=(v_1,v_2,\dots,v_n)\in (\F_{q^2}^*)^n$.
For all $f(x)\in \F_{q^2}[x]$ with form $f(x)=\sum_{i=0}^{k-2}f_ix^i+f_{k-1}(x^{k-1}+\eta x^k)$.
Set $h(x)=\sum_{i=0}^{k-2}f_i^qx^i+f_{k-1}^q(x^{k-1}+\eta^q x^k)$.
Since $\alpha_i^q=a\alpha_i+b$,
then
\[\begin{split}
	f(\alpha_i)^q &=\sum_{i=0}^{k-2}f_i^q(\as_i^q)^i+f_{k-1}^q((\as_i^q)^{k-1}+\eta^q (\as_i^q)^k)\\
            &=h(a\alpha_i+b).\\
\end{split}\]
Set $g(x)=\lambda h(ax+b)$.
To make $g(x)$ has form $g(x)=\sum_{i=0}^{k-2}g_ix^{i}+g_{k-1}(x^{k-1}-\frac{\eta}{1+s(\aaa)\eta}x^k)$,
by analyzing the coefficient of $x^{k-1}$ and $x^k$ on both sides, then
$$\frac{\eta^q a}{1+k\eta^q b}=-\frac{\eta}{1+s(\aaa)\eta}.$$
It follows that $(a s(\aaa)+kb)\eta^q+a\eta^{q-1}+1=0$.
Thus we have that if $\eta\in \F_{q^2}^*$ satisfies
$$(a s(\aaa)+kb)\eta^q+a\eta^{q-1}+1=0,$$ then $g(x)$ has the desired form.
Therefore, there exists a $g(x)$ such that  $v_i^{q+1}f(\as_i)^q=L_{\aaa}(\as_i)^{-1}g(\as_i)$, $1\leq i\leq n$.
By Lemma \ref{lem +TGRS}, $TGRS_{k,\eta}^{+}(\aaa,\bm{v})$ is an $[n,k]_{q^2}$ Hermitian self-dual GRS code.
 This completes the proof.
\end{proof}

\begin{remark}
Theorem \ref{th +TGRS} generalizes the previous results and leads to new $q^2$-ary Hermitian self-dual $(+)$-TGRS codes.
\begin{itemize}
\item If $a=1$ and $b=(\theta^q-\theta)a_l$, where $\theta$ is a primitive element of $\F_{q^2}$ and $a_l\in \F_q$, then \cite[Theorem 5]{RefJ (2023) Guo H-TGRS} is a special case of Theorem \ref{th +TGRS}.

\item If $a=\beta_m^{q-1}$ and $b=(1-\beta_m^{q-1})a_l$, where $\theta$ is a primitive element of $\F_{q^2}$, $\beta_m=\theta^m,\ 1\leq m\leq q$ and $a_l\in \F_q$, then \cite[Theorem 6]{RefJ (2023) Guo H-TGRS} is a special case of Theorem \ref{th +TGRS}.
 \item Comparing to \cite[Theorems 5 and 6]{RefJ (2023) Guo H-TGRS}, we consider the case $2\mid q$ and $s(\aaa)=0$.
 Next we give an example which cannot be obtained from previous results in the literature.
\end{itemize}
\end{remark}

\begin{example}
Let $q=2^m$ and $\theta$ be a primitive $q$-th root of unity.
If $a=1$ and $b=0$, then $\F_q=S_{1,0}$.
Let $\aaa=(0,1,\theta,\theta^2,\dots,\theta^{q-2})\in \F_q^q= S_{1,0}^q$, then $s(\aaa)=0$. By Theorem \ref{th +TGRS}, for any $\eta\in \F_q^*$,
there exists a vector $\bm{v}\in (\F_{q^2}^*)^q$ such that $TGRS_{\frac{q}{2},\eta}^{+}(\aaa,\bm{v})$ is a $[q,\frac{q}{2}]_{q^2}$ Hermitian self-dual $(+)$-TGRS code.
\end{example}

By Lemma \ref{lem +TGRS MDS} and Theorem \ref{th +TGRS}, we can directly obtain the following corollary.

\begin{corollary}
For given $a\in \F_{q^2}^*$ and $b\in \F_{q^2}$.
Assume that integers $n=2k$ satisfy that there exist $n$ distinct elements $\alpha_1,\alpha_2,\dots,\alpha_n$ such that $\alpha_i\in S_{a,b}$ for $1\leq i\leq n$.
Let $\aaa=(\alpha_1,\alpha_2,\dots,\alpha_n)\in S_{a,b}^n$ and $[n]=\{1,2,\dots,n\}$,
then the following two assertions hold.
\begin{itemize}
\item[(1)] If $\eta\in \F_{q^2}^*$ such that $(a s(\aaa)+kb)\eta^q+a\eta^{q-1}+1=0$ and for any $k$-subset $I\subseteq [n]$,
$\eta \sum_{i\in I}\as_i\neq -1$ holds,
then there exists a vector $\bm{v}=(v_1,v_2,\dots,v_n)\in (\F_{q^2}^*)^n$ such that $TGRS_{k,\eta}^{+}(\aaa,\bm{v})$ is an $[n,k]_{q^2}$ Hermitian self-dual MDS code.
\item[(2)] If $\eta\in \F_{q^2}^*$ such that $(a s(\aaa)+kb)\eta^q+a\eta^{q-1}+1=0$ and
there exists $k$-subset $I\subseteq [n]$ such that $\eta \sum_{i\in I}\as_i=-1$,
then there exists a vector $\bm{v}=(v_1,v_2,\dots,v_n)\in (\F_{q^2}^*)^n$ such that $TGRS_{k,\eta}^{+}(\aaa,\bm{v})$ is an $[n,k]_{q^2}$ Hermitian self-dual NMDS code.
\end{itemize}

\end{corollary}

\subsubsection{Hermitian self-dual $(\ast)$-TGRS codes}

The following lemma shows that $(\ast)$-TGRS codes is MDS or NMDS.

\begin{lemma}\label{lem *TGRS MDS}
(\cite{RefJ (2017) Bee TGRS})
Let $k\leq n\leq q$,
$\aaa=(\as_1,\as_2,\dots,\as_n)\in \F_{q}^n$, with $\as_i\neq \as_j\ (i\neq j)$, $\bm{v}=(v_1,v_2,\dots,v_n)\in (\F_{q}^*)^n$ and $\eta\in \F_q^*$.
Let $[n]=\{1,2,\dots,n\}$,
then the following two assertions hold.
\begin{itemize}
\item[(1)]
$TGRS_{k,\eta}^{\ast}(\aaa,\bm{v})$ is MDS if and only if for any $k$-subset $I\subseteq [n]$, it holds
$\eta \prod_{i\in I}\as_i\neq (-1)^k$.
\item[(2)]
$TGRS_{k,\eta}^{\ast}(\aaa,\bm{v})$ is NMDS if and only if
there exists $k$-subset $I\subseteq [n]$ such that $\eta \prod_{i\in I}\as_i=(-1)^k$.
\end{itemize}
\end{lemma}

The following lemma gives the non-GRS properties of $(\ast)$-TGRS codes.

\begin{lemma}\label{lem *TGRS non-GRS}
(\cite{RefJ (2024) Zhu*})
For $3\leq k\leq n-3$, $TGRS_{k,\eta}^{\ast}(\aaa,\bm{v})$ is not GRS or EGRS.
\end{lemma}

By Theorem \ref{th TGRS not exist}, if $n\geq 8$ and $0\in A_{\aaa}$, there exists no $q^2$-ary Hermitian self-dual $(\ast)$-TGRS code.
Then we need only consider the case $0\notin A_{\aaa}$.
In \cite{RefJ (2024) Zhu*}, the authors give the Euclidean dual codes of $(\ast)$-TGRS codes.

\begin{lemma}
(\cite{RefJ (2024) Zhu*})
Let $\aaa=(\as_1,\as_2,\dots,\as_n)\in (\F_{q^2}^*)^n$ with $\as_i\neq \as_j\ (i\neq j)$, $\bm{v}=(v_1,v_2,\dots,v_n)\in (\F_{q^2}^*)^n$ and $\eta\in \F_{q^2}^*$.
Then
$$TGRS_{k,\eta}^{\ast}(\aaa,\bm{v})^{\bot_E}=TGRS_{n-k,\eta^{-1}p(\aaa)^{-1}}^{\ast}(\aaa,\aaa^{-1}\star \bm{v}^{-1}\star L_{\aaa}^{-1}).$$
\end{lemma}

Next, we give a sufficient and necessary condition for $(\ast)$-TGRS codes to be Hermitian self-dual.

\begin{lemma}\label{lem H-dual *TGRS}
Let $n$ be even, $\aaa=(\as_1,\as_2,\dots,\as_n)\in (\F_{q^2}^*)^n$, with $\as_i\neq \as_j\ (i\neq j)$ and $\bm{v}=(v_1,v_2,\dots,v_n)\in (\F_{q^2}^*)^n$,
then $TGRS_{\frac{n}{2},\eta}^{*}(\aaa,\bm{v})$ over $\F_{q^2}$ is Hermitian self-dual if and only if
there exists a polynomial
$g(x)=\sum_{i=1}^{k-1}g_ix^{i}+g_{0}(1+\eta^{-1}p(\aaa)^{-1}x^k)$,
$g_i\in \F_{q^2}$, $0\leq i\leq k-1$ such that
$$v_i^{q+1}f(\as_i)^q=\as_i^{-1} L_{\aaa}(\as_i)^{-1}g(\as_i),\ 1\leq i\leq n,$$
for all $f(x)\in \mathcal{V}_{\frac{n}{2},1,0,\eta,q^2}$.
\end{lemma}

\begin{proof}
Let $TGRS_{\frac{n}{2},\eta}^{*}(\aaa,\bm{v})$ have a generator matrix $G_{\frac{n}{2},\eta}^{*}(\aaa,\bm{v})$.
Note that
$TGRS_{\frac{n}{2},\eta}^{*}(\aaa,\bm{v})$ over $\F_{q^2}$ is Hermitian self-dual
if and only if for any codeword
$\bm{c}=(v_1f(\as_1),v_2f(\as_2),\dots,v_nf(\as_n))$ of $TGRS_{\frac{n}{2},\eta}^{*}(\aaa,\bm{v})$,
we have
\[\begin{split}
&\bm{c}^q\cdot G_{\frac{n}{2},\eta}^{*}(\aaa,\bm{v})^T \\
            =&(v_1^{q+1}f^q(\as_1),v_2^{q+1}f^q(\as_2),\dots,v_n^{q+1}f^q(\as_n))\cdot G_{\frac{n}{2},\eta}^{*}(\aaa,\bm{1})^T\\
            =&\bm{0}\\
           \Leftrightarrow& (v_1^{q+1}f^q(\as_1),v_2^{q+1}f^q(\as_2),\dots,v_n^{q+1}f^q(\as_n))\in TGRS_{\frac{n}{2},\eta}^{\ast}(\aaa,\bm{1})^{\bot_E} \\
           \Leftrightarrow& (v_1^{q+1}f^q(\as_1),v_2^{q+1}f^q(\as_2),\dots,v_n^{q+1}f^q(\as_n))\in TGRS_{\frac{n}{2},\eta^{-1}p(\aaa)^{-1}}^{\ast}(\aaa,\aaa^{-1}\star L_{\aaa}^{-1}).\\
\end{split}\]
Then we know that
$TGRS_{\frac{n}{2},\eta}^{*}(\aaa,\bm{v})$ is  Hermitian self-dual if and only if
there exists a polynomial $g(x)=\sum_{i=1}^{k-1}g_ix^{i}+g_{0}(1+\eta^{-1}p(\aaa)^{-1}x^k)$,
$g_i\in \F_{q^2}$, $0\leq i\leq k-1$ such that
$$v_i^{q+1}f(\as_i)^q=\as_i^{-1} L_{\aaa}(\as_i)^{-1}g(\as_i),\ 1\leq i\leq n,$$
for all $f(x)\in \mathcal{V}_{\frac{n}{2},1,0,\eta,q^2}$.
 This completes the proof.
\end{proof}

By Lemma \ref{lem H-dual *TGRS}, we give the following construction of Hermitian self-dual $(\ast)$-TGRS codes.

\begin{theorem}\label{th *TGRS}
Let $n=2k$, $\aaa=(\alpha_1,\alpha_2,\dots,\alpha_n)\in (\F_q^*)^n$, where $\alpha_1,\alpha_2,\dots,\alpha_n$
are distinct elements of $\F_q^*$. If $\eta\in \F_{q^2}^*$ such that $\eta^{q+1}=p(\aaa)^{-1}$,
 then there exists a vector $\bm{v}=(v_1,v_2,\dots,v_n)\in (\F_{q^2}^*)^n$ such that $TGRS_{k,\eta}^{\ast}(\aaa,\bm{v})$ is an $[n,k]_{q^2}$ Hermitian self-dual $(\ast)$-TGRS code.
\end{theorem}

\begin{proof}
Since $\as_i\in \F_q^*$, for $1\leq i\leq n$, then $\as_i^{-1} L_{\aaa}(\as_i)^{-1}\in \F_q^*$.
Then there exist $v_i\in \F_{q^2}^*$ such that $v_i^{q+1}=\as_i^{-1} L_{\aaa}(\as_i)^{-1}$ for all $1\leq i\leq n$.
Set $\bm{v}=(v_1,v_2,\dots,v_n)\in (\F_{q^2}^*)^n$.
For all $f(x)\in \F_{q^2}[x]$ with form $f(x)=\sum_{i=1}^{k-1}f_ix^i+f_{0}(1+\eta x^k)$.
Set $g(x)=\sum_{i=1}^{k-1}f_i^qx^i+f_{0}^q(1+\eta^qx^k)$.
Note that $\alpha_i^q=\alpha_i$,
then
\[\begin{split}
	f^q(\alpha_i) &=\sum_{i=1}^{k-1}f_i^q(\as_i^q)^i+f_{0}^q(1+\eta^q (\as_i^q)^k)\\
            &=g(\alpha_i).\\
\end{split}\]
To make $g(x)$ has form
$g(x)=\sum_{i=1}^{k-1}g_ix^{i}+g_{0}(1+\eta^{-1}p(\aaa)^{-1}x^k)$,
by analyzing the coefficient of $x^{0}$ and $x^k$ on both sides, then
$$\eta^q=\eta^{-1}p(\aaa)^{-1}.$$
It follows that $\eta^{q+1}=p(\aaa)^{-1}$.
Note that $p(\aaa)^{-1}\in \F_q^*$, then there exists $\eta\in \F_{q^2}^*$ such that $\eta^{q+1}=p(\aaa)^{-1}$.
Therefore, there exists a $g(x)$ such that  $v_i^{q+1}f(\as_i)^q=\as_i^{-1}L_{\aaa}(\as_i)^{-1}g(\as_i)$, $1\leq i\leq n$.
By Lemma \ref{lem H-dual *TGRS}, $TGRS_{k,\eta}^{\ast}(\aaa,\bm{v})$ is an $[n,k]_{q^2}$ Hermitian self-dual $(\ast)$-TGRS code.
 This completes the proof.
\end{proof}

\begin{example}\label{example *TGRS}
Let $n=2k$ and $2\leq n\leq q-1$,
then there exists an $[n,k]_{q^2}$ Hermitian self-dual $(\ast)$-TGRS code.
\end{example}

\begin{remark}
Example \ref{example *TGRS} states that there exist $q^2$-ary Hermitian self-dual $(\ast)$-TGRS codes for each even length $n\leq q-1$.
\end{remark}

Similarly, by Lemma \ref{lem *TGRS MDS} and \ref{lem *TGRS non-GRS} and Theorem \ref{th *TGRS}, we can directly obtain the following corollary.

\begin{corollary}
Let $n=2k\geq 6$, $\aaa=(\alpha_1,\alpha_2,\dots,\alpha_n)\in (\F_q^*)^n$, where $\alpha_1,\alpha_2,\dots,\alpha_n$
are distinct elements of $\F_q^*$.
Let $[n]=\{1,2,\dots,n\}$,
then the following two assertions hold.
\begin{itemize}
\item[(1)] If $\eta\in \F_{q^2}^*$ such that $\eta^{q+1}=p(\aaa)^{-1}$ and for any $k$-subset $I\subseteq [n]$,
$\eta \prod_{i\in I}\as_i\neq (-1)^k$ holds,
then there exists a vector $\bm{v}=(v_1,v_2,\dots,v_n)\in (\F_{q^2}^*)^n$ such that $TGRS_{k,\eta}^{\ast}(\aaa,\bm{v})$ is an $[n,k]_{q^2}$ Hermitian self-dual non-GRS MDS code.
\item[(2)] If $\eta\in \F_{q^2}^*$ such that $\eta^{q+1}=p(\aaa)^{-1}$ and
there exists $k$-subset $I\subseteq [n]$ such that $\eta \prod_{i\in I}\as_i=(-1)^k$,
then there exists a vector $\bm{v}=(v_1,v_2,\dots,v_n)\in (\F_{q^2}^*)^n$ such that $TGRS_{k,\eta}^{\ast}(\aaa,\bm{v})$ is an $[n,k]_{q^2}$ Hermitian self-dual NMDS code.
\end{itemize}

\end{corollary}

\section{Conclusion}\label{sec5}

In this paper, we give a sufficient and necessary condition for the dual extended code $\underline{\C}[\bm{u}]$ of $\C$ to be $e$-Galois self-orthogonal (see Theorem \ref{th galois 1}).
From this, a new systematic approach is proposed to prove the existence of $e$-Galois self-dual codes.
By this method, we prove that $e$-Galois self-dual (extended) GRS codes of length $n>\min\{p^e+1,p^{m-e}+1\}$ do not
exist, where $1\leq e\leq m-1$ (see Theorem \ref{th GRS not exist}).
Moreover, based on the non-GRS properties of TGRS codes, we show that in many cases $e$-Galois self-dual TGRS and ETGRS codes do not exist
(see Theorems \ref{th TGRS not exist} and \ref{th ETGRS not exist}).
Since Galois self-dual codes are generalizations of Euclidean and Hermitian self-dual codes,
we can also obtain the existence of some Euclidean and Hermitian self-dual GRS and TGRS codes directly.
Furthermore, we construct several new classes of Hermitian self-dual $(+)$-TGRS and $(\ast)$-TGRS codes
(see Theorems \ref{th +TGRS} and \ref{th *TGRS}).

\section*{Acknowledgments}

This research is supported by the National Natural Science Foundation of China under Grant Nos. 12171134 and U21A20428.

\section*{Data availability}

All data generated or analyzed during this study are included in this paper.

\section*{Conflict of Interest}

The authors declare that there is no possible conflict of interest.

\end{document}